\documentclass[aps,prb,amsmath,amssymb,twocolumn,floatfix,10pt,showpacs,superscriptaddress]{revtex4-1}

\usepackage{graphicx}
\usepackage{color}
\usepackage{epstopdf}

\begin{document}

\title{Correlations and entanglement in quantum critical bilayer and necklace XY models}
\author{Johannes Helmes}
\affiliation{Institut f\"ur Theoretische Festk\"orperphysik, JARA-FIT and JARA-HPC, RWTH Aachen University, 52056 Aachen, Germany}
\affiliation{Institut f\"ur Theoretische Physik, Universit\"at zu K\"oln, 50937 K\"oln, Germany}
\author{Stefan Wessel}
\affiliation{Institut f\"ur Theoretische Festk\"orperphysik, JARA-FIT and JARA-HPC, RWTH Aachen University, 52056 Aachen, Germany}
\date{\today}

\begin{abstract}
We analyze the critical properties and the entanglement scaling at the quantum critical points of the spin-half XY model on the 
two-dimensional square-lattice bilayer and necklace lattice,
based on quantum Monte Carlo simulations on finite tori and for different subregion shapes. For both models, the finite-size scaling 
of the transverse staggered spin structure factor is found in accord 
with a quantum critical point described by the two-component, three-dimensional $\phi^4$-theory. The second R\'enyi entanglement entropy in the absence of corners along the
subsystem boundary exhibits area-law scaling in both models, with an area-law prefactor of $0.0674(7)$ 
[$0.0664(4)$] for the bilayer [necklace] model, respectively. Furthermore, the presence of $90^{\circ}$ 
corners leads to an additive logarithmic term in both models. We estimate 
a contribution of $-0.010(2)$ [$-0.009(2)$] due to each $90^{\circ}$ corner 
to the logarithmic correction for the bilayer [necklace] model,  and compare our findings 
to recent numerical linked cluster calculations and series expansion results on related models.  
\end{abstract}

\maketitle

\section{Introduction}

The study of the entanglement in quantum many-body systems has lead to new insights into the structure 
of strongly correlated  quantum states. 
One interesting direction of current research in this respect is the identification of universal
contributions to the scaling of the bipartite entanglement entropy in quantum many-body systems. 
For the special case of one-dimensional quantum critical states, described by a conformal field theory, 
it is well known, for example, that the entanglement entropy $S$ asymptotically scales as $S=c \ln(l)/3$ with the subregion size $l$~\cite{holzey94,calabrese04}. 
Here, the central charge $c$ provides a universal number, which furthermore also relates to the number of degrees of freedom, 
e.g., for bosonic free theories.  
For higher-dimensional quantum systems, similar universal contributions to the entanglement entropy scaling require one to 
consider corrections beyond the leading scaling form, which in most generic cases is set by the ``area-law'' scaling 
of the entanglement entropy $S$ with the size of the boundary that separates a subregion from the rest of the system~\cite{bombelli86,srednicki93,eisert08}. 
For two-dimensional systems, on which we shall focus here, this leading asymptotic behavior reads $S=a l$, in terms of the length $l$ of the subregion's boundary, 
with a non-universal area-law prefactor $a$ that depends explicitly on microscopic details of the system under consideration. 

Several distinct contributions to the subleading scaling of $S$ with $l$ have been considered recently, related, e.g., to topological order~\cite{kitaev06,levin06} or the presence of Goldstone modes~\cite{kallin11,metlitski11,kulchytskyy15}. 
Here, we consider in 
particular 
the case of a quantum critical many-body system. For this case, a growing body of evidence has been obtained from both numerical studies of various quantum many-body lattice models, as well as field-theoretical calculations in the continuum limit,  that an isolated corner in the subregion boundary adds a logarithmic contribution, i.e.,  a term $c_\mathrm{c}(\theta) \ln(l)$, to the bipartite entanglement entropy~\cite{ardonne04,fradkin06,casini07,hirata07,singh12,inglis13,kallin13,kallin14,helmes14,devakul14a,stoudenmire14,bueno15,laflorencie15}. In order to obtain a lattice regularized version of such corner terms without introducing further  lattice artifacts, boundaries with  $\theta=90^{\circ}$ corners may be most conveniently considered in numerical studies on 
finite square lattices. 
On a more quantitative level, these calculations provide strong evidence that the prefactor $c_\mathrm{c}(\theta)$ for a corner with an opening angle $\theta$ scales to a high precision at least in leading order  proportional to the number of field components $N$ of the critical O($N$) theory, with a prefactor that appears to be a universal function of $\theta$ for all critical  
$\phi^4$-theory cases considered thus far. 

\begin{figure}[!t]
\includegraphics[width=0.9\linewidth]{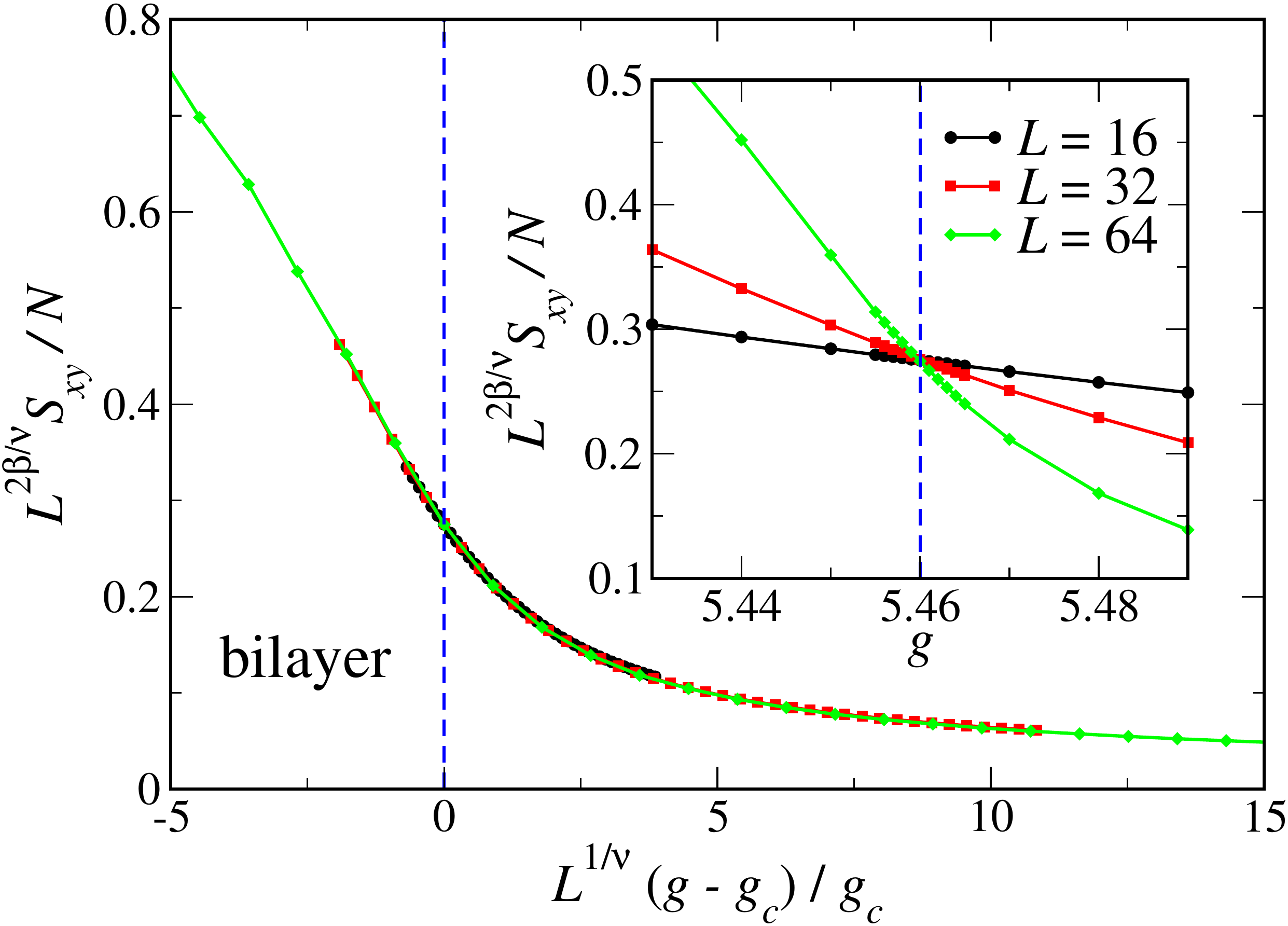}
\includegraphics[width=0.9\linewidth]{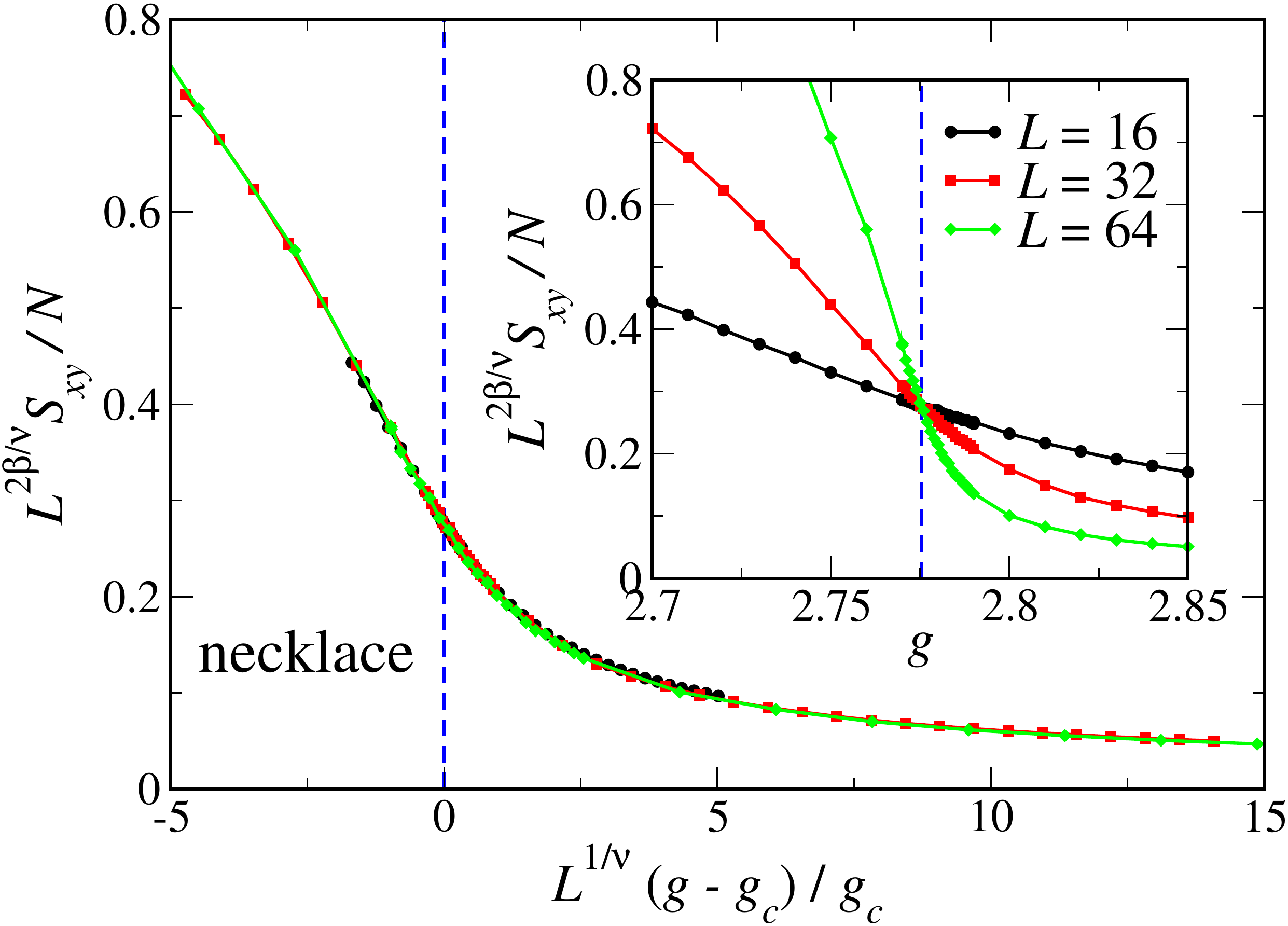}
\caption{(Color online) Data collapse plot of the transverse staggered spin structure factor $S_{xy}$ within the quantum critical region of the bilayer (top panel) and the necklace (lower panel) mode. The insets show the rescaled transverse staggered spin structure factor as a function of $g$. Dashed lines indicate the quantum critical points. Error bars in this figure are below the symbol size.}
\label{fig:sf}
\end{figure}

While for the cases of $N=1$ and $N=3$, support for this observation has been  provided by series and numerical linked cluster expansions, as well as by quantum Monte Carlo simulations,~\cite{singh12,inglis13,kallin13,kallin14,helmes14,devakul14a} the case of $N=2$ still needs to be addressed by quantum Monte Carlo studies. Here, we complement recent results from 
series expansion and numerical linked-cluster studies~\cite{devakul14a,stoudenmire14} on several two-dimensional lattice models with O(2) symmetry by a quantum Monte Carlo estimate of 
the $\theta=90^{\circ}$ corner term. In particular, we consider the case of the quantum critical spin-half XY model on the square-lattice bilayer, which provides a basic model for probing the entanglement properties at a quantum critical point with an $O(2)$ critical theory. For this model, we provide estimates for both the leading area-law prefactor $a$ as well as the corner term $c_\mathrm{c}(90^{\circ})$, based on the second ($\alpha=2$) R\'enyi~\cite{renyi61} entropy-based bipartite entanglement measure
\begin{equation}
 S_\alpha(A)=\frac{1}{1-\alpha} \ln\mathrm{Tr}[(\rho_A)^\alpha], 
\end{equation}
where $\rho_A$ denotes the reduced density matrix of the subregion (denoted $A$). 
In addition, we also consider the quantum critical spin-half XY model a the square necklace lattice (or incomplete bilayer), which, in contrast to the bilayer model, has a finite spin exchange interaction within only one of the two layers.  
We introduce both these models and locate their quantum critical points based directly on their magnetic properties in Sec.~II. This also allows us to confirm the expected critical exponents from a finite-size scaling analysis. 
In Sec.~III, we then present our computational scheme and  extract the entanglement entropy scaling coefficients for $S_2$ for both these models. 
These values are finally discussed in comparison to previous results in Sec.~IV. 

\section{Quantum Critical Point}

In the following, we consider the spin-half XY model on a square-lattice bilayer, described by the Hamiltonian
\begin{eqnarray}
 H&=&J\sum_{\langle i,j\rangle} \sum_{l=1}^2\left( {S}^x_ {i,l}{S}^x_ {j,l} +{S}^y_ {i,l}{S}^y_ {j,l}  \right)\nonumber \\
&&\:+\: J_{\perp}\sum_i \left( {S}^x_ {i,1} {S}^x_ {i,2}+{S}^y_ {i,1} {S}^y_ {i,2} \right),
\end{eqnarray}
where $i$ denotes the $i$th unit cell containing two spin-half degrees of freedom (associated to the two layers, $l=1,2$), and $J$ ($J_\perp$) the intralayer (inter-layer) exchange interaction.

In addition, we consider the spin-half XY model on a  square necklace lattice (or incomplete bilayer), described 
by the Hamiltonian
\begin{eqnarray}
 H&=&J\sum_{\langle i,j\rangle} \left( {S}^x_ {i,1}{S}^x_ {j,1} +{S}^y_ {i,1}{S}^y_ {j,1}  \right)\nonumber \\
&&\:+\: J_{\perp}\sum_i \left( {S}^x_ {i,1} {S}^x_ {i,2}+{S}^y_ {i,1} {S}^y_ {i,2} \right),
\end{eqnarray}
in which the inter-layer coupling has been turned to zero in one of the layers. This corresponds to a square-lattice spin system 
with a local impurity spin attached to each lattice site, i.e., a Kondo necklacelike model with XY exchange interactions.
The SU(2)-symmetric version of this model has been considered previously using quantum Monte Carlo methods, both with respect to  ground state 
properties~\cite{wang06} and at finite temperatures~\cite{brenig06}.
In the following, we 
denote by $g=J_\perp/J$ the ratio of the interlayer to the intralayer exchange interactions for both models considered here. 

Both models exhibit a quantum phase transition at a critical coupling ratio $g=g_c$ between a low-$g$ phase with long-range
transverse antiferromagnetic order to a large-$g$ quantum disordered phase. The  
structure factor corresponding to the order parameter for this transition is given in terms of the transverse  spin correlations as
\begin{equation}
 S_{xy}=\frac{1}{N}\sum_{i,j=1}^N \epsilon_i\epsilon_j \langle S^x_i S^x_j + S^y_i S^y_j \rangle.
\end{equation}
Here, the summations are performed over all spins on the finite lattice, 
where $\epsilon_i=\pm 1$, depending on the sublattice to which spin $i$ belongs on the bipartite lattices. We consider, in particular, finite lattices of linear extent $L$, containing $N=2L^2$ spins, employing periodic boundary conditions in both lattice directions. For $g<g_c$, $S_{xy}/N$ extrapolates to a finite value in the thermodynamic limit. 
In Ref.~\onlinecite{stoudenmire14}, an estimate of $g_c=5.460(1)$ was obtained for the bilayer model from 
quantum Monte Carlo simulations combined with a finite-size scaling analysis of the spin stiffness 
$\rho_S$, employing the fact that at the quantum critical point, $\rho_S$ scales linearly with $L$, 
reflecting a dynamical critical exponent $z=1$. Here, we confirm this value of $g_c$ from calculations of the structure factor $S_{xy}$, which allows us to verify explicitly also that the quantum critical behavior is indeed in accord with the expected three-dimensional $O(2)$ universality class. For this purpose, we performed quantum Monte Carlo simulations employing the stochastic series expansion approach~\cite{sandvik99} for finite lattices with periodic boundary conditions,  scaling the inverse temperature $1/T=4L$ with linear system size in order to probe ground state correlations. Near the quantum critical point, the finite-size data exhibit  conventional finite-size scaling behavior,
\begin{equation}
 S_{xy}/N=L^{-2\beta/\nu} G(L^{1/\nu}(g-g_c)/g_c),
\end{equation}
with a scaling function $G$ and  critical exponents $\beta$ and $\nu$, that for the three-dimensional $O(2)$ universality class take on the values $\beta=0.3486(1)$ and 
$\nu=0.6717(1)$, respectively~\cite{campostrini06}. When performing simulations within the critical region, the above scaling form implies a common crossing point of the various finite-size values of the rescaled structure factor $L^{2\beta/\nu}S_{xy}/N$  at the quantum critical point, i.e., for $g=g_c$. This fact allows us to locate the quantum critical point as shown in the upper inset of Fig.~1. We obtain a value of $g_c=5.460(1)$, in agreement with the previous estimate, based on the spin stiffness~\cite{stoudenmire14}. Furthermore, the finite-size data exhibit an excellent data collapse, confirming again the extracted value of $g_c$ as well as the employed critical exponents of the three-dimensional $O(2)$ universality class (cf. the main upper panel of Fig.~1).

We are not aware of previous estimates of the location of the quantum critical point for the XY necklace model, and thus performed a similar finite-size analysis as for the XY bilayer model, with the resulting data collapse and crossing plots shown in the lower panel of Fig.~1. From our analysis, we obtain as estimate of $g_c=2.7755(5)$ for the XY necklace model, and again find excellent accord of the numerical data with the anticipated 
three-dimensional $O(2)$ universality class. 

\section{Entanglement Scaling}

After having established the value of the quantum critical coupling ratios and confirming the three-dimensional $O(2)$ universality 
class 
of the quantum phase transitions on both lattice geometries, 
we next performed quantum Monte Carlo simulations to extract the scaling properties of the second R\'enyi entropy $S_2$ 
at these quantum critical points. In order to separate the logarithmic contribution arising from the corners in the subregion boundary, 
we considered in each case two differently shaped subregion types, similarly to our procedure for the SU(2)-symmetric Heisenberg bilayer 
case~\cite{helmes14}. First, we consider a bipartition of the toroidal simulation cell into two equally sized cylindrical strips, 
both of size $L/2 \times L$. The circumference of the subregion boundary in this case equals $l=2L$. The striplike subregions  
exhibit smooth boundaries without any corners. In order to introduce corners into the finite 
discrete-lattice 
subregion boundary in a controlled, scalable way, we considered in addition the case of a square subregion of size $L/2 \times L/2$, 
this way introducing four $90^{\circ}$ corners along the subregion boundary. Upon increasing the linear system size $L$, 
we thus scaled the subregion size in both cases such that the aspect ratio remained at a constant value. 
Any contribution to $S_2$ that depends 
only on the aspect ratio thus reduces for both considered subregions to a $l$-independent, constant term. We refer to 
Refs.~\cite{inglis13,laflorencie15} 
for a discussion of various proposed functional forms of such aspect ratio contributions to the entanglement entropy. 
For both subregion types, we then calculated $S_2$ for various boundary lengths $l$, employing the extended ensemble sampling 
approach, based on the replica trick~\cite{hastings10,melko10} within the stochastic series expansion quantum Monte Carlo representation of 
Ref~\onlinecite{humeniuk12}. We furthermore used the "increment trick"~\cite{hastings10,humeniuk12} to successively obtain the entanglement entropy upon 
growing the subregion for an efficient sampling. 

\begin{figure}[!t]
\includegraphics[width=\linewidth]{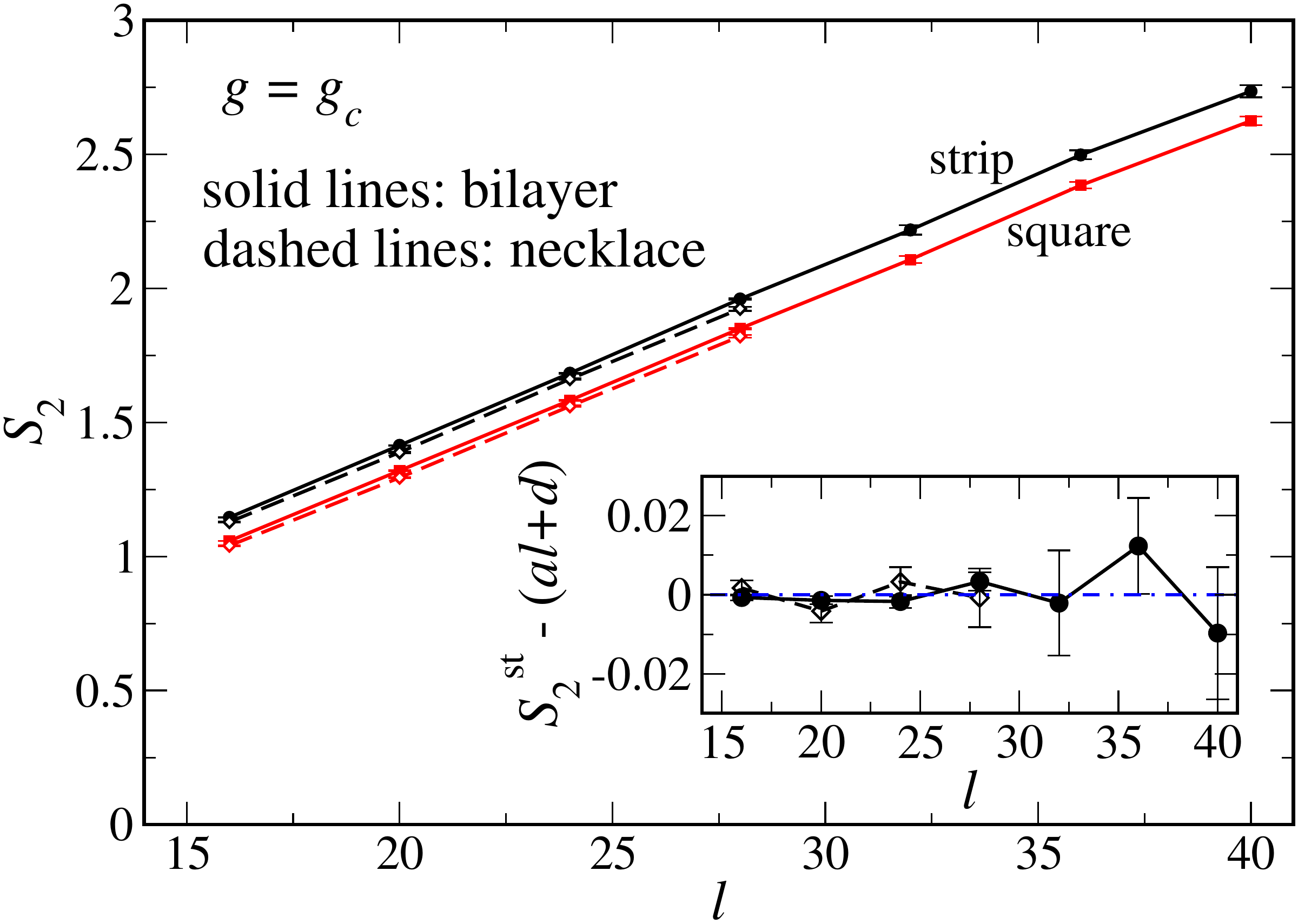}
\caption{(Color online)
Second R\'enyi entropy $S_2$ as a function of the subregion boundary length $l$ at the critical coupling ratio $g=g_c$ for strip and square shaped subregions.
The inset shows the residuals to a linear fit $al+d$ of  $S^\mathrm{st}_2$ for strip (st) shaped subregions boundary length $l$.
}
\label{fig:re}
\end{figure}

The results for the boundary length dependence of the entanglement entropy $S_2$ at the quantum critical point 
for both 
subregion types are shown in Fig.~2 for both the bilayer and the necklace lattice. We find that in both cases, $S_2$ exhibits a dominant area-law scaling.
We first considered the case of the XY bilayer model, for which we considered system sizes from $L=8$ to $L=20$, 
corresponding to boundary lenghs between $l=16$ and $l=40$, 
respectively.
As shown in the 
insets of Fig.~2, the strip subregion data is well accounted for by the area-law. 
From fitting the finite-size data for the strip subregions to a linear scaling form,
we obtain an area-law scaling coefficient 
of $a=0.0674(7)$ for the bilayer model.
One notices also that the statistical uncertainty 
in our numerical 
data increases with increasing subregion size. This is due to the propagation of errors while employing the incremental 
procedure to access larger subregion sizes. We thus concentrated our computational resources towards 
the lower four system sizes, where accurate results for $S_2$ are more readily accessible.  
Therefore, for the necklace model systems we considered sizes from $L=8$ to $L=14$, corresponding to $l=16$ and $l=28$, respectively.
In our simulations we found it not feasible to extent the considered system sizes to larger values of $L$, due to enhanced  statistical uncertainties
for the necklace model, presumably due to the lower values of the critical coupling strength $g_c$ in that model. 
From fitting the finite-size data for the strip subregions to a linear scaling from, we obtain $a=0.0664(4)$ for the quantum critical necklace model. While this is not significantly different from the above quoted 
value for the quantum critial bilayer model, no universal meaning is associated to the area-law prefactor, as it depends on microscopic details. 

\begin{figure}[!t]
\includegraphics[width=\linewidth]{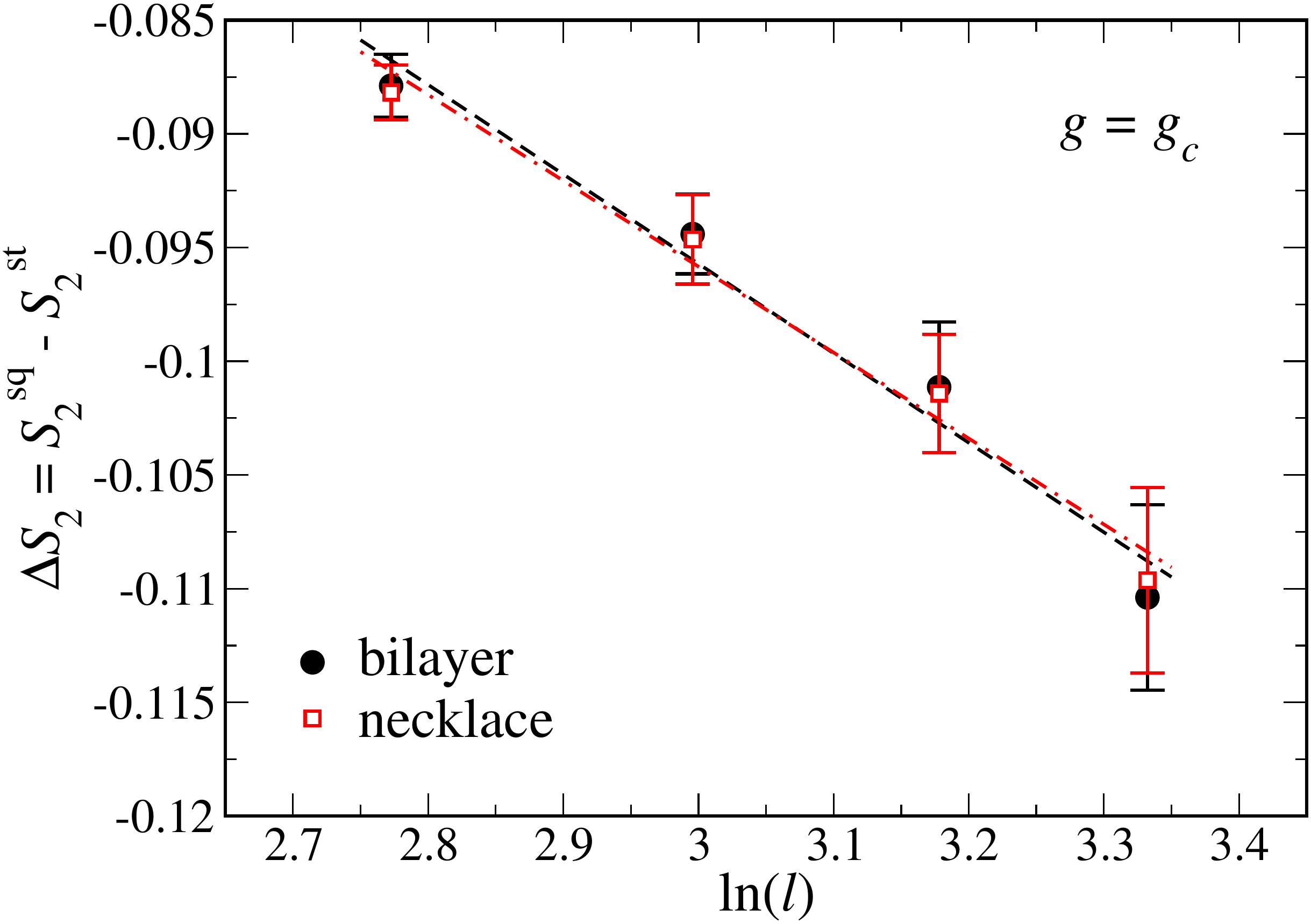}
\caption{(Color online) Difference between the second R\'enyi entropy $\Delta S_2=S_2^\mathrm{sq}-S_2^\mathrm{st}$ for square (sq) 
and 
strip (st)  subregions as a function of the subregion boundary length $l$ (shown on a log-linear scale) at the critical coupling ratio $g=g_c$, along with the best fit line (dashed line). 
}
\label{fig:Svsl}
\end{figure}

For square subregions, we expect, in addition to the leading area law a logarithmic contribution 
to the entanglement scaling, due to the presence of the four 
corners along the subregion boundary. However, 
based on the finite-size data accessible to our numerical study, it is not feasible to reliably extract this logarithmic 
contribution from a direct fit to the square subregion data. In fact, we found that for square subregions on the restricted 
available $l$-range, 
the 
$S_2$ data may be fit within the statistical uncertainty also to a linear $l$-scaling. In order to estimate therefore the 
prefactor 
of the logarithmic term due to the presence of the four corners in the square subregion, we instead followed the procedure from 
Ref.~\onlinecite{helmes14}, and considered directly the difference $\Delta S_2=S_2^\mathrm{sq}-S_2^\mathrm{st}$  between the  
second 
R\'enyi entropies for the square (sq) and strip (st) subregions. This quantity is directly accessible  within the quantum Monte Carlo simulations,
using the fact that 
$\Delta S_2 = S_2^{sq} - S_2^{st} = -\ln ({Z[A^\mathrm{sq},2,T]}/{Z^2}) + \ln
(Z[A^\mathrm{st},2,T]/{Z^2}) = -\ln ({Z[A^\mathrm{sq},2,T]}/{Z[A^\mathrm{st},2,T]})$,
where $Z$ denotes the thermal partition function of the total lattice system, and $Z[A,2,T]$ the replica-trick ensemble partition function~\cite{humeniuk12} for 
subregions $A=A^\mathrm{st}$ or $A^\mathrm{sq}$, respectively. 
Our data for $\Delta S_2$ for both the bilayer and necklace models are shown in Fig.~3 as a function of $\ln(l)$ for the 
lower four system sizes. 
While the 
data still exhibit statistical uncertainties, 
they fit well for both models to a linear dependence 
$\Delta S_2=4 \times c_\mathrm{c}(90^\circ)\ln(l)+\Delta d$. 
The resulting value of 
$4\times c_\mathrm{c}(90^\circ)=-0.039(7)$ for the bilayer model implies a contribution to the logarithmic entanglement scaling of $S_2$ of 
$c_\mathrm{c}(90^\circ)=-0.010(2)$ for each $90^\circ$ corner for the bilayer model, while for the 
necklace lattice model we obtain
values of $4\times c_\mathrm{c}(90^\circ)=-0.037(7)$, and $c_\mathrm{c}(90^\circ)=-0.009(2)$, 
respectively.
Both values of $c_\mathrm{c}(90^\circ)$ compare rather well to each other, in accord with the expectation that the corner contribution to the entanglement scaling 
exhibits a universal character.
It should be noted that the above error bars on the fitting parameters account 
for the statistical uncertainties in our quantum Monte Carlo data, 
but do not reflect possible systematic deviations due to further 
subleading finite-size corrections in the $l$-scaling of the entanglement entropy 
(cf. also our previous discussion in Ref.~\onlinecite{helmes14}).

\section{Discussion}
Based on quantum Monte Carlo calculations of the transverse spin correlations, we identified the quantum critical points 
of the spin-half XY model on the square-lattice bilayer and necklace lattice. 
Our result for the critical coupling ratio for the bilayer model agrees with a previous estimate based on the spin 
stiffness, and 
furthermore exhibits finite-size scaling in accord with a three-dimensional O(2) critical $\phi^4$-theory universality class, as does our finite-size data for the necklace 
model.
At the quantum critical points, we extracted the scaling prefactors of the dominant area law in the second R\'enyi bipartite entanglement entropy, as well as an additional 
logarithmic term from $90^\circ$ corners in the subregion boundary, with consistent values for the two 
different models.
Our results for 
the scaling prefactor furthermore compare  well to a recent estimate of $c_\mathrm{c}(90^\circ)=-0.0111(1)$ from numerical linked-cluster 
calculations~\cite{stoudenmire14}, obtained for various two-dimensional O(2) quantum critical spin systems.
They are also consistent within the statistical uncertainty with the values of $-0.0125(6)$ and $-0.0127(13)$, reported from series expansions~\cite{devakul14a}. Given that the employed
computational approaches access this contribution through a rather different analysis [finite systems (here) vs.  thermodynamic-limit linked-cluster calculations],  
we consider our results to add further support for a possible universal character of such corner terms in the bipartite entanglement measure. For the future, it will be important to apply similar methods also to quantum phase transitions that reside outside the conventional $\phi^4$-theory framework, e.g. the XY${}^*$ universality class~\cite{sentil02,grover11}, accessible in specifically designed quantum many-body systems~\cite{isakov12}.\\

%\acknowledgements

We acknowledge discussions with P. Br\"ocker,  T. Grover, D. J. Luitz, R. G. Melko, R. R. P. Singh, and E. M. Stoudenmire. Financial support by the Deutsche Forschungsgemeinschaft under Grant WE 3649/3-1 is acknowledged, as well as the allocation of CPU time within
JARA-HPC at JSC J\"ulich and at RWTH Aachen University. J.H. thankfully acknowledges support from the Bonn-Cologne Graduate School for Physics and Astronomy. 
S.W. thanks the KITP Santa Barbara for hospitality during the program ``Entanglement in Strongly-Correlated Quantum Matter``.
This research was supported in part by the National Science Foundation under Grant No. NSF PHY11-25915.

%
% ----bib sorted according to citation order----
%
\end{document}